\documentstyle[aps,multicol,epsf]{revtex}
\begin{document}
\draft
\title{The Pinning Paths of an Elastic Interface}

\author{Hern\'an A. Makse,\footnote{present address: Cavendish
Laboratory, University of Cambridge, Cambridge CB3 0HE, UK} Sergey Buldyrev, 
Heiko Leschhorn,\footnote{present address: Theoretische Physik III,
Heinrich-Heine-Universit\"at D\"usseldorf, D-40225 D\"usseldorf,
Germany}
and H. Eugene Stanley}

\address{Center for Polymer Studies and Dept. of Physics \\
 Boston University, Boston, Massachusetts 02215}

\date{\today}   

\maketitle

\begin{abstract}
We introduce a model describing the paths that pin 
an elastic interface moving in a disordered medium.
We find that the scaling properties of these ``elastic pinning paths'' (EPP) 
are different from paths embedded on a directed percolation cluster,
which are known to pin the interface of the ``directed percolation
depinning'' class of surface growth models.
The EPP are characterized 
by a roughness exponent $\alpha=1.25$, intermediate
between that of the free inertial process ($\alpha=3/2$) and the
diode-resistor problem on a Cayley tree ($\alpha=1$). 
We also calculate
numerically the mean cluster size and the cluster size distribution for
the EPP.
\end{abstract}


\begin{multicols}{2}
\narrowtext

The problem of interface roughening in the presence of quenched
disorder is a topic of recent interest, 
due to its importance as a 
paradigm in condensed matter physics and due to the broad range of
applications \cite{review}.  In a typical case, the interface moves in a
$(d+1)$-dimensional disordered medium driven by a homogeneous force
$F$.
At small forces, the interface is pinned by the impurities of the
medium, while 
the interface undergoes a depinning transition at a critical force
$F_c$, and
for $F>F_c$ the interface moves with a nonzero velocity.
The spatial fluctuations of the interface are
characterized by the scaling of the saturated interface width $W_{sat}(L)$ with
the system size, 
\begin{equation}
W_{sat}(L) \sim L^\alpha,
\label{wit}
\end{equation}
where $\alpha$ is the roughness
exponent.

It has been proposed \cite{Feigelman,bruinsma} that
the depinning transition 
can be described by the following equation of motion for the
interface height $y(\vec{x},t)$
\begin{equation}
{\partial\over\partial t}y(\vec{x},t)=\nabla^2 y+\eta(\vec{x},y)+F,
\label{qew}
\end{equation}
where $\vec{x}$ is the
$d$-dimensional coordinate parallel to the interface. 
The first term on the right
hand side of (\ref{qew}) represents the surface tension favoring 
a smooth interface, and we say that the interface is elastic. 
The second term is a random field that mimics the
quenched disorder of the medium and is assumed to have zero mean and 
short-range correlations.  

The universality class corresponding 
to Eq. (\ref{qew}) is called quenched Edwards-Wilkinson
(QEW), because  (\ref{qew})
is similar to the
Edwards-Wilkinson equation \cite{ew}. The difference,
which changes the behavior of (\ref{qew}) drastically,
is the
presence of spatially dependent 
quenched disorder $\eta(\vec{x},y)$ instead of time-dependent 
shot noise $\eta(\vec{x},t)$. 
Numerical studies \cite{lesch,makse,PacMas}
of the depinning transition 
yield a roughness exponent $\alpha\simeq1.25$
in $d=1$ ($(1+1)$ dimensions) and $\alpha \simeq 0.75$ for
$d=2$ ($(2+1)$ dimensions).
These values are lower than the results of perturbation theory 
\cite{Feigelman}, 
which yields $\alpha = 3/2$ in $d=1$ and $\alpha = 1$
for $d=2$. On the other hand, the numerical values are 
significantly higher than the prediction of a functional 
renormalization group treatment which gives $\alpha \simeq 1$ 
and $\alpha \simeq 2/3$ for $d=1$ and $d=2$, respectively 
\cite{NSTL}. 

The relevance of directed percolation to interface depinning 
has been established for 
a different class of models, called
{\it directed percolation depinning} models \cite{qkpz}, 
which are in the same universality
class as Eq. (\ref{qew}) when a 
Kardar-Parisi-Zhang (KPZ) \cite{kpz} term $\lambda (\nabla y)^2$ is
included.  
In these models, the interface is pinned by paths on a 
directed percolation cluster of pinning sites \cite{qkpz}.
Thus, the scaling
properties of the interface at the depinning transition in $(1+1)$
dimensions can be obtained by a mapping onto directed percolation (DP)
\cite{dp,ds}, from which a roughness exponent $\alpha \simeq 0.63$ is
obtained.

In this paper, we consider the paths which 
pin the interface for the QEW universality class for the case $d=1$. We term
these paths elastic pinning paths (EPP).
We apply the concepts of directed percolation to investigate 
the scaling properties of the EPP.
Our numerical results provide 
an independent check for the anomalous roughness exponent $\alpha \simeq
1.25$ obtained with the models of the QEW universality class.
We also consider two known random walk models, 
which yield $\alpha = 3/2$ and $\alpha = 1$, respectively. 
By comparing the EPP
to these random walks we obtain some insight why the 
roughness exponent of Eq. (\ref{qew}) lies in the interval 
$1 < \alpha < 3/2$.      

To motivate the definition of the EPP, we first consider 
a discrete solid-on-solid model corresponding to Eq. (\ref{qew})
\cite{lesch}. The interface position $h_i$ 
is defined on a square lattice of lateral size $L$.
We assign to each site on the lattice a random number $\eta _{i,h}$
which can have two values, $\eta_{i,h}=1$ (unblocked cell) 
with probability $p$, and
$\eta_{i,h}=-1$ (blocked cell) with probability $1-p$.  A local force is
defined by 
\begin{equation}
\label{force}
f_i \equiv  h_{i+1} +  h_{i-1} - 2 h_i + \eta_{i,h_i}.
\end{equation}
At time $t=0$ the interface is flat, and at a given time the height of the
$i$-th column is increased by one if the local force $f_i$ is positive.

At a critical value of the probability $p=p_c$, the interface is pinned
by one of the pinning paths. According to the dynamical
rules, a pinned interface satisfies $f_i\leq 0$ for all $i$. We define
the increments of a given path 
as $\Delta_i \equiv h_{i}-h_{i-1}$, so according to
Eq. (\ref{force}) a spanning path stopping the QEW interface satisfies
\begin{equation}
\label{condition}
\Delta_{i+1} \leq \Delta_{i} -  \eta_{i,h}.
\end{equation}
By induction, one can show that in addition to the condition
(\ref{condition}), a lower bound for $\Delta_{i+1} - \Delta_{i}$ holds
at every time step of the interface evolution
\begin{equation}
\label{condition2}
\Delta_{i+1} -  \Delta_{i} \geq -2.
\end{equation}

Equations (\ref{condition}) and (\ref{condition2}) define the possible
pinning paths.
Note that paths in the same cell $(i,h)$ can have different increments 
$ \Delta _i$.
We start at $i=1$
with $h_1=0$ and initial increment $\Delta_1=0$. 
The paths in column $i+1$ are updated according to the following three rules:
 
{\it \bf Rule (i):} If the cell $(i,h)$ is blocked $(\eta_{i,h}=-1)$, then the
path is splitted into four paths, where   
the positions at $i+1$ are $h_{i+1} = h_i + \Delta_i + 1$, 
$h_{i+1} = h_i + \Delta_i$,
$h_{i+1} = h_i + \Delta_i - 1$, and $h_{i+1} = h_i + \Delta_i - 2$

{\it \bf Rule (ii):} If the cell $(i,h)$ is unblocked,
$\eta_{i,h}=1$,we have:
$h_{i+1} = h_i + \Delta_i - 1$, and $h_{i+1} = h_i + \Delta_i - 2$.

{\it \bf Rule (iii):} The path stops when $h_i\leq 0$.

\begin{figure}
\narrowtext
\centerline{
%$(a)$
\vbox{ \hbox{\epsfxsize=7cm \epsfbox{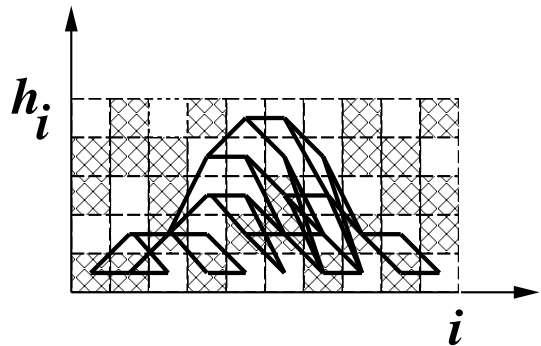} } }
 }
\caption{Example of one cluster of pinning paths generated by 
the proposed rules. The sites in a square lattice are unblocked ($\eta
=1$, shown in white) with probability $p$ and blocked ($\eta =-1$, shown
as cross-hatched) with probability $1-p$.  The path starts at $x_1 = 0$
with $h_1 = 0$ and $\Delta_1 =0$ and we apply the rules {\it \bf (ib)},
{\it \bf (iia)}, and {\it \bf (iii)}.  }
\label{path}
\end{figure}

After moving to the new cell the increment
$\Delta_{i+1}$ is updated and the rules are applied again. 
Rules {\it \bf (i)} and {\it \bf (ii)} are the implemetation of Eq.
(\ref{condition}). Rule {\it
\bf (iii)} is motivated by the fact that if the path deviates too much
in the downward direction, it would not have a chance to block the
growth since, in a system with periodic boundary conditions, the path
should return to the same point where it starts.  In Fig. \ref{path} we
show a typical set (``cluster'') 
of directed paths. The paths are characterized by
large local slopes which is the main feature of the QEW
interface at the depinning transition.

The scaling properties of the directed paths require two
characteristic lengths, $\xi_\parallel$ and $\xi_\perp$, the
correlation length parallel to and perpendicular to the preferred
direction of the paths \cite{dp-exponents}.  
The correlation lengths diverge at the
critical concentration of pinning centers $p_c$ as
\begin{equation}
\xi_\parallel(p) \sim |p-p_c|^{-\nu_\parallel} \qquad\qquad
\xi_\perp(p) \sim |p-p_c|^{-\nu_\perp},
\label{connectedness}
\end{equation}
where $\nu_\parallel$ and $\nu_\perp$ are two different universal
exponents due to the anisotropy given by the preferred direction of the
paths.

A cluster consisting of pinning paths is defined by all paths generated by the
Rules {\bf (i-iii)} for one realization of the disorder.
Let us denote by $s$ the number of sites in a given cluster.
The cluster size
distribution $n_s(p)$, defined as the average number of clusters of $s$
sites per lattice site, shows a power-law behavior at $p_c$. For
$p<p_c$ only finite clusters are present, so there exists an effective
cutoff for the cluster size, $s_0\sim |p-p_c|^{-1/\sigma}$. 
The cluster size distribution has the scaling form
$ n_s(p) \sim s^{-\tau} ~ g(s/s_0)$,
where $g(x)$ is a scaling
function that decreases faster than a power-law for $x\gg 1$.  The mean
cluster size $\langle s \rangle$ also diverges at $p_c$ as a power-law
$
\langle s\rangle  \sim |p-p_c|^{-\gamma}$,
with $\gamma = (2 - \tau ) / \sigma$.

The scaling relations presented so far are valid for infinite
lattices.  
Finite size scaling considerations allow us to
write \cite{qkpz}
$W(L)\sim  \xi_\perp \sim \xi_\parallel ^{\nu_\perp/\nu_\parallel} \sim
L^{\nu_\perp/\nu_\parallel}$.
Hence from (\ref{wit})
\begin{equation}
\alpha = \nu_\perp/\nu_\parallel.
\label{mapping2} 
\end{equation}

\begin{figure}
\narrowtext
\centerline{
\vbox{ \hbox{\epsfxsize=7cm \epsfbox{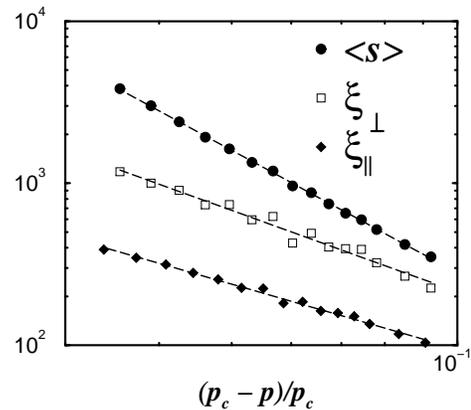} } }
 }
\vspace{.5cm}
\caption{
Log-log plots of the 
mean cluster size $<s>$ (shifted for clarity), the 
correlation lengths $\xi_\perp$ and $\xi_\parallel$ 
in the perpendicular 
and parallel
directions
as a function of the reduced probability $(p_c -
p)/p_c$. Simulations are averaged over $10^7$ clusters.}
\label{results}
\end{figure}

In our simulations we compute all the exponents characterizing the
scaling behavior of the EPP.  Results are shown in
Fig. \ref{results}. We calculate the correlation length exponents
and find $\nu_\parallel \simeq 1.33$ and $\nu_\perp\simeq1.67$ (see
Table \ref{table}).  Using (\ref{mapping2}) we find $\alpha \simeq 1.25$
in agreement with the exponent found for the QEW interface
\cite{lesch,makse,PacMas}.  The study of the mean cluster size yields an
exponent $\gamma \simeq 2.43$, while the exponent of the cluster
size distribution is $\tau\simeq 1.43$.

Next we calculate the scaling of the mean square fluctuations or
``width'' of the paths defined as 
$\langle h_i^2\rangle^{1/2}$ as a
function of the parallel coordinate $i$ calculated at $p_c$. We find
$\langle
h_i^2\rangle^{1/2} \sim i^\alpha$ with 
$\alpha\simeq 1.27$, in agreement with the exponent
$\alpha=\nu_\perp/\nu_\|=1.25$ that we find using (\ref{mapping2}) and
our numerical results for $\nu _\perp $ and $\nu _\parallel$.

We also compute the mean square fluctuations of the increments
$\Delta_i$.
We find $\langle \Delta_i^2 \rangle ^{1/2} \sim i^{\alpha'}$ where
$\alpha'=\alpha-1$ because $h_i$ is the integral of $\Delta_i$ along the path,
so by integration the exponent $\alpha$ is $1+\alpha'$.

It is interesting to note that Rules {\it \bf (i)} and {\it \bf (ii)}
can be modified without changing the universality class. Indeed, instead
of four choices for $\Delta_{i+1}$ in Rule {\it \bf (i)} and two choices
for $\Delta_{i+1}$ in {\it \bf (ii)}, we can set $\Delta_{i+1} =
\Delta_{i} + k$, $k=-1, 0, 1$ (Rule {\it \bf (ia)}), and $\Delta_{i+1} =
\Delta_{i} -1 $, (Rule {\it \bf (iia)}), or even $\Delta_{i+1} =
\Delta_{i} + k$, with $k=-1, 1$ (Rule {\it \bf (ib)}).  These changes
are analogous to the change of coordination number in directed
percolation which change the value of the critical threshold but do not
change the universality class. The results in Table \ref{table} 
correspond to the most extensive simulations which are performed with
the Rules {\it \bf (ib)} and {\it \bf (iia)}.

Next, we discuss the relation of the EPP to two known universality
classes ---which can be considered as random walk models---and to
which the EPP can be modified by changing the interaction 
with the disorder.
For the first universality class we assume that the noise is determined
by the position $(i,\Delta_i)$ in the increment space,
instead 
by the real space position $(i,h_i)$ for the EPP.
Since the average span of $\Delta _i \sim i^{\alpha '}$ is much smaller
than the average span of $h_i \sim i^{\alpha}$,
many of the EPP that were treated differently (since they have
different $h_i$ coordinate) become indistinguishable, since many of them
can have the same value of $\Delta_i$, so that
the number of different paths decreases significantly compared to the
EPP.  

\begin{figure}
\centerline{
\vbox{ \hbox{\epsfxsize=7cm  \epsfbox{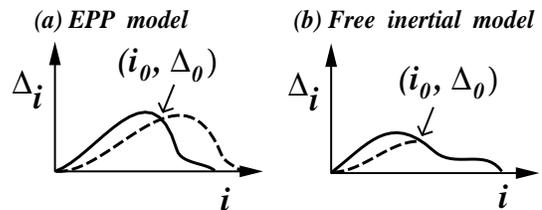} } }
 }
\narrowtext
\caption{Schematic illustration of the top-most  trajectory in the
$(i,\Delta_i)$ plane for $(a)$ the EPP, and $(b)$ the modification
of the EPP where the noise is chosen according to the position of
the path in the $(i,\Delta_i)$ plane (free inertial process).  In both
figures we plot two paths (solid and dashed lines) which intersect at
the point $(i_0,\Delta_0)$. In the EPP case the paths can cross each
other, because the noise is determined by the position in real space
$(i,h)$. Thus, also paths in real space can cross each other.
In the free inertial process, however,
if two paths intersect at $(i_0,\Delta_0)$ then they continue
together. Therefore, paths in real space cannot cross each other.}
%Then the top-most path in the EDP model is composed of several different
%paths, and the top-most trajectory in the free inertial process corresponds
%to the trajectory of only one single path that is indeed a random walk
%in the $(x,\Delta_i)$ plane.}
\label{cross}
\end{figure}

The span of the cluster is determined by the top-most
trajectory in real space which is in turn represented by 
the top-most trajectory in the increment space $(i,\Delta_i)$.
Since the noise
is chosen according to the position in the $(i,\Delta_i)$ space, 
two paths which arrive to the
same point $(i_0,\Delta_0)$ remain together because also
the noise is the same.
Thus, two given paths cannot cross each other in the
plane $(i,\Delta_i)$ (see Fig. \ref{cross}) and the 
top-most trajectory in the increment space is composed by a unique and
well-defined path.
This path is a simple random walk in the
($i,\Delta_i$) plane, which has a certain probability of going up and down
depending on the noise. Therefore the average span scales as
$\Delta _i\sim i^{\alpha'}$ with $\alpha'=1/2$. With 
$\alpha = \alpha '+1$ we get 
$\alpha =3/2 > \alpha_{EPP} = 1.25$.
This is in agreement with the fact that the standard deviation
of the free inertial process \cite{massoliver,havlin}
increases as $i^{3/2}$.

The critical probability, $p_c$, now corresponds to the unbiased random
walk, and can be readily computed analytically for each of the
modification of the Rules {\it \bf (i)} and {\it \bf (ii)}. The critical
probabilities are always larger that the corresponding values in the EPP
case, being $p_c = 1/2$ for the rules {\it \bf (ib)} and {\it \bf
(iia)}.  The exponent $\tau$ now is related to the probability of the
first return to the origin after some number of steps $i$. For the free
inertial process this probability decays as $i^{-5/4}$ \cite{goldman},
which differs from the random walk result $i^{-3/2}$. This indicates
that the top-most trajectory in the $(i,\Delta_i)$ plane 
determines only the scaling of the span of the clusters, but
other properties, such as the probability of first return to the origin, are
not determined by the top-most trajectory.

For the EPP, where the noise is chosen from the position in
real space ($i,h$), the top-most trajectory in the $(i,\Delta_i)$ plane
still determines the scaling properties of the span on the clusters. 
However, in this case, the top-most 
trajectory in the increment space can be 
composed by parts  of several different paths (Fig.
\ref{cross}).
This is so because
two paths can cross each other in the
($i,\Delta_i$) plane because the noise is determined by the 
position in the ($i,h$) plane which can be different for the two paths
at the crossing point $(i_0,\Delta_0)$ (Fig. \ref{cross}).
This is the main difference between the EPP and
the free inertial process.

The second universality class to which the EPP reduces
is obtained by
defining the value of the noise for each path independently.
Even for two paths that meet at a point ($i,h_i$) with the {\it same} 
$\Delta_i$, the values of the noise are chosen as 
independent of each other.
Since now all
paths are independent and there are no loops,
this model can be
exactly mapped---with respect to $\Delta_i$ and $i$---to the diode
resistor Cayley tree problem solved in \cite{sergey3}.
The $i$ coordinate is identified with the time coordinate of the
Cayley tree cluster which, in turn, corresponds to the chemical or minimum
path in the longitudinal hyperplane of the
Cayley tree cluster (see \cite{shlomo}), and has $\nu_\parallel = 1/2$
\cite{sergey3}. The coordinate $\Delta_i$ is the remaining
transversal coordinate of the Cayley tree cluster 
and has  $\nu_\perp = 0$\cite{sergey3}.
Then, we find $\alpha'=0$, and  $\alpha=\alpha'+1 =
1 < \alpha_{EPP} = 1.25$. We also 
find numerically $\alpha'=0$. 
The critical probability $p_c$ in this case can be computed
analytically, and it decreases in comparison with the EPP case. For
rules {\it \bf (ib)} and {\it \bf (iia)} $p_c = 1/4$.

What can we learn from the comparison of the EPP to the two
discussed random walk models?
Since the EPP can cross each other in the $(i, \Delta_i$) plane, 
there is no unique top-most path
but a top-most trajectory consists of several paths.  
By being at some point not the top-most
path, a given path can ``optimize'' its way through the
randomness in the sense that it visits more blocked cells.
Thus, it can stay alive longer ($h_i > 0$)  compared to 
the free inertial process where the unique top-most path
determines the scaling properties. 
This means that the EPP has a smaller 
roughness exponent than the free inertial value of $3/2$.

The other random walk model, where the noise of different paths 
is always independent, has a roughness exponent $\alpha = 1$ which
is even lower than that of the EPP. In order to understand 
the reason, we note that the paths have at each point many more 
possibilities than the EPP. Thus, it is not surprising that 
there are always paths which stay alive even for large $i$.
It seems reasonable that the scaling properties are dominated
by these paths, causing a smaller roughness exponent.

To summarize, we introduce the EPP model to describe 
the paths that pin 
an elastic interface moving in a disordered medium,
and we find that it is  in a
universality class different from directed percolation paths that are
responsible for pinning of the class of models of the directed
percolation depinning universality class. The
critical paths that are characterized by an exponent $\alpha_{EPP} =
1.25$, describe the scaling properties of an elastic interface pinned by
the quenched disorder of the medium.  The roughness exponent of the EPP
lies between that of the free inertial process with
$\alpha=1.5$ \cite{massoliver}, and that of the diode resistor Cayley
tree problem with $\alpha=1$ \cite{sergey3}.  The comparison 
of the EPP with these two random walk models sheds some
light on the value of $\alpha=1.25$.  

We would like to thank D. Ertas and H. K. Janssen for useful
discussions. H. L. acknowledges support from the DFG. The CPS is
supported by NSF.

\begin{table}
\narrowtext
\caption{ Critical exponents for the EPP presented in
this paper along with the results for the elastic interface in the
QEW universality class, and the DP universality class. We find
$p_c=0.282\pm0.001$ for the EPP.}
\begin{tabular}{|l|c|c|c|}
     & EPP           & QEW (interface)           & DP \\
\tableline
$~\alpha$      & $1.26\pm0.03$ & $1.25\pm0.01$ \cite{lesch}
& $0.63\pm0.04$ \cite{qkpz} \\
$~\nu_\parallel$& $1.33\pm0.04$ & $1.35\pm0.04$ \cite{makse}
& $1.73\pm0.01$ \cite{dp-exponents} \\
$~\nu_\perp$   & $1.67\pm0.04$ &  $1.68\pm0.04$ & $1.09\pm0.01$
\cite{dp-exponents}   \\
$~\tau$        & $1.43\pm0.02$ &  ---          & $1.28\pm0.02$
\cite{dp-exponents}   \\
$~\gamma$      & $2.43\pm0.05$ &  ---          & $2.28\pm0.01$
\cite{dp-exponents}   \\
\end{tabular}
\label{table}
\end{table}

\end{multicols}

\end{document}